\documentclass[12pt,a4paper]{article}
\usepackage{psfrag}
\usepackage{graphics}
\usepackage{amssymb,epsfig,amsmath,euscript,array}

\makeatletter
\@addtoreset{equation}{section}
\makeatother
\renewcommand{\theequation}{\thesection.\arabic{equation}}

\newcommand{\startappendix}{
\setcounter{section}{0}
\renewcommand{\thesection}{\Alph{section}}
\renewcommand{\theequation}{\Alph{section}.\arabic{equation}}}

\newcommand{\Appendix}[1]{
\refstepcounter{section}
\begin{flushleft}
{\Large\bf Appendix \thesection: #1}
\end{flushleft}}

\newcounter{multieqs}

\newcommand{\be}{\begin{equation}}
\newcommand{\ee}{\end{equation}}

\newcommand{\bm}[1]{\mbox{\boldmath $#1$}}

\def\bd{\begin{document}}
\def\ed{\end{document}}
\def\nn{\nonumber}
\def\bea{\begin{eqnarray}}
\def\eea{\end{eqnarray}}
\let\bm=\bibitem
\let\la=\label

\newcommand{\EQ}[1]{\begin{equation} #1 \end{equation}}
\newcommand{\AL}[1]{\begin{subequations}\begin{align} #1 \end{align}\end{subequations}}
\newcommand{\SP}[1]{\begin{equation}\begin{split} #1 \end{split}\end{equation}}
\newcommand{\ALAT}[2]{\begin{subequations}\begin{alignat}{#1} #2 \end{alignat}\end{subequations}}
\def\beqa{\begin{eqnarray}}
\def\eeqa{\end{eqnarray}}
\def\beq{\begin{equation}}
\def\eeq{\end{equation}}

\def\hf{{\textstyle \frac{1}{2}}}
\def\wbar{\bar w}
\def\mubar{\bar\mu}
\def\abar{\bar a}
\def\sigmabar{\bar\sigma}
\def\etabar{\bar\eta}
\def\zetabar{\bar\zeta}
\def\mubar{\bar\mu}
\def\nubar{\bar\nu}
\def\N{{\cal N}}
\def\sst{\scriptscriptstyle}
\def\thetabar{\bar\theta}
\def\Tr{{\rm Tr}}
\def\one{\mbox{1 \kern-.59em {\rm l}}}
 \def\Nh{\hat{N}}

\newlength{\myVSpace}
\newcommand\xstrut{\raisebox{-.5\myVSpace}
  {\rule{0pt}{\myVSpace}}%
}

\def\a{\alpha}      \def\da{{\dot\alpha}}
\def\b{\beta}       \def\db{{\dot\beta}}
\def\c{\gamma}  \def\G{\Gamma}  \def\cdt{\dot\gamma}
\def\d{\delta}  \def\D{\Delta}  \def\ddt{\dot\delta}
\def\e{\epsilon}        \def\vare{\varepsilon}
\def\f{\phi}    \def\F{\Phi}    \def\vvf{\f}
\def\h{\eta}
\def\k{\kappa}
\def\l{\lambda} \def\L{\Lambda}
\def\m{\mu} \def\n{\nu}
\def\o{\omega}
\def\p{\pi} \def\P{\Pi}
\def\r{\rho}
\def\s{\sigma}  \def\S{\Sigma}
\def\t{\tau}
\def\th{\theta} \def\Th{\Theta} \def\vth{\vartheta}
\def\X{\Xeta}
\def\z{\zeta}

\def\cA{{\cal A}} \def\cB{{\cal B}} \def\cC{{\cal C}}
\def\cD{{\cal D}} \def\cE{{\cal E}} \def\cF{{\cal F}}
\def\cG{{\cal G}} \def\cH{{\cal H}} \def\cI{{\cal I}}
\def\cJ{{\cal J}} \def\cK{{\cal K}} \def\cL{{\cal L}}
\def\cM{{\cal M}} \def\cN{{\cal N}} \def\cO{{\cal O}}
\def\cP{{\cal P}} \def\cQ{{\cal Q}} \def\cR{{\cal R}}
\def\cS{{\cal S}} \def\cT{{\cal T}} \def\cU{{\cal U}}
\def\cV{{\cal V}} \def\cW{{\cal W}} \def\cX{{\cal X}}
\def\cY{{\cal Y}} \def\cZ{{\cal Z}}

\def\ua{\underline{\alpha}}
\def\ub{\underline{\phantom{\alpha}}\!\!\!\beta}
\def\uc{\underline{\phantom{\alpha}}\!\!\!\gamma}
\def\um{\underline{\mu}}
\def\ud{\underline\delta}
\def\ue{\underline\epsilon}
\def\una{\underline a}\def\unA{\underline A}
\def\unb{\underline b}\def\unB{\underline B}
\def\unc{\underline c}\def\unC{\underline C}
\def\und{\underline d}\def\unD{\underline D}
\def\une{\underline e}\def\unE{\underline E}
\def\unf{\underline{\phantom{e}}\!\!\!\! f}\def\unF{\underline F}
\def\unm{\underline m}\def\unM{\underline M}
\def\unn{\underline n}\def\unN{\underline N}
\def\unp{\underline{\phantom{a}}\!\!\! p}\def\unP{\underline P}
\def\unq{\underline{\phantom{a}}\!\!\! q}
\def\unQ{\underline{\phantom{A}}\!\!\!\! Q}
\def\unH{\underline{H}}

\def\As {{A \hspace{-6.4pt} \slash}\;}
\def\bs {{b \hspace{-6.4pt} \slash}\;}
\def\Ds {{D \hspace{-6.4pt} \slash}\;}
\def\ds {{\del \hspace{-6.4pt} \slash}\;}
\def\ss {{\s \hspace{-6.4pt} \slash}\;}
\def\ks {{ k \hspace{-6.4pt} \slash}\;}
\def\ps {{p \hspace{-6.4pt} \slash}\;}
\def\pas {{{p_1} \hspace{-6.4pt} \slash}\;}
\def\pbs {{{p_2} \hspace{-6.4pt} \slash}\;}

\def\Fh{\hat{F}}
\def\Vh{\hat{V}}
\def\Xh{\hat{X}}
\def\ah{\hat{a}}
\def\xh{\hat{x}}
\def\yh{\hat{y}}
\def\ph{\hat{p}}
\def\xih{\hat{\xi}}

\def\psit{\tilde{\psi}}
\def\Psit{\tilde{\Psi}}
\def\tht{\tilde{\th}}

\def\At{\tilde{A}}
\def\Qt{\tilde{Q}}
\def\Rt{\tilde{R}}
\def\Nt{\tilde{N}}

\def\at{\tilde{a}}
\def\st{\tilde{s}}
\def\ft{\tilde{f}}
\def\pt{\tilde{p}}
\def\qt{\tilde{q}}
\def\vt{\tilde{v}}
\def\nt{\tilde{n}}

\def\delb{\bar{\partial}}
\def\bz{\bar{z}}
\def\bD{\bar{D}}
\def\bB{\bar{B}}

\def\bk{{\bf k}}
\def\bl{{\bf l}}
\def\bp{{\bf p}}
\def\bq{{\bf q}}
\def\br{{\bf r}}
\def\bx{{\bf x}}
\def\by{{\bf y}}
\def\bR{{\bf R}}
\def\bV{{\bf V}}

\def\d{\delta}\def\D{\Delta}\def\ddt{\dot\delta}

\def\pa{\partial} \def\del{\partial}
\def\xx{\times}
\def\uno{\mbox{1 \kern-.59em {\rm l}}}

\def\trp{^{\top}}
\def\inv{^{-1}}
\def\dag{{^{\dagger}}}
\def\pr{^{\prime}}

\def\rar{\rightarrow}
\def\lar{\leftarrow}
\def\lrar{\leftrightarrow}

\newcommand{\0}{\,\!}      
\def\one{1\!\!1\,\,}
\def\im{\imath}
\def\jm{\jmath}

\newcommand{\tr}{\mbox{tr}}
\newcommand{\slsh}[1]{/ \!\!\!\! #1}

\def\vac{|0\rangle}
\def\lvac{\langle 0|}

\def\hlf{\frac{1}{2}}
\def\ove#1{\frac{1}{#1}}

\def\Box{\square}
\def\ZZ{\mathbb{Z}}
\def\CC#1{({\bf #1})}
\def\bcomment#1{}
\def\bfhat#1{{\bf \hat{#1}}}
\def\VEV#1{\left\langle #1\right\rangle}
\def\vev#1{\langle{#1}\rangle}

\newcommand{\ex}[1]{{\rm e}^{#1}} \def\ii{{\rm i}}

\def\rr{{\rm r}} \def\rs{{\rm s}}\def\rv{{\rm v}}
\def\ri{{\rm i}}\def\rj{{\rm j}}

\newcommand{\lrbrk}[1]{\left(#1\right)}
\newcommand{\sfrac}[2]{{\textstyle\frac{#1}{#2}}}

\font\mybb=msbm10 at 12pt
\def\bb#1{\hbox{\mybb#1}}

\font\myBB=msbm10 at 18pt
\def\BB#1{\hbox{\myBB#1}}

\begin{document}

\hfill{ hep-th/0406178}

\vspace{20pt}

\begin{center}

{\Large \bf  Noncommutative Standard Modelling}

\vspace{30pt}

{\bf Valentin V. Khoze and Jonathan Levell}

{\small \em
Centre for Particle Theory,
Department of Physics and IPPP,\\
University of Durham, Durham, DH1 3LE, UK
}

\vspace{10pt}

{\sffamily \tt
valya.khoze@durham.ac.uk,
jonathan.levell@durham.ac.uk}

\vspace{30pt}
{\bf Abstract}

\end{center}
We present a noncommutative gauge theory that has the ordinary Standard Model
as its low-energy limit. The model is based on the gauge group
$U(4) \times U(3) \times U(2)$ and is constructed to satisfy the key
requirements imposed by noncommutativity: the UV/IR mixing effects,
restrictions on representations and charges of matter fields, and
the cancellation of noncommutative gauge anomalies. At energies
well below the noncommutative mass scale our model flows to the
commutative Standard Model plus additional free $U(1)$ degrees of freedom
which are decoupled due to the UV/IR mixing. Our model also predicts
the values of the hypercharges of the Standard Model fields.
\setcounter{page}{0}
\thispagestyle{empty}
\newpage


\section{Introduction}

One of the most novel and intriguing aspects of
noncommutative gauge theories\footnote{For reviews
of noncommutative gauge theories and an extensive list of references
see \cite{SW,dn,szabo}.}
is the UV/IR mixing in which the physics of high-energy
degrees of freedom affects the physics at low energies \cite{Minwalla, MST}.
Gauge theories on spaces with noncommuting coordinates,
\EQ{
[x^\mu,x^\nu]=i\theta^{\mu\nu} \ , }
arise naturally as low-energy effective theories from string theory and D-branes,
but they
are also known to be extremely restrictive and difficult to use in particle
physics model building due to a number of field-theoretical constraints
imposed by noncommutativity:
\begin{enumerate}
\item{} the UV/IR mixing \cite{Minwalla,MST} and decoupling of
$U(1)$ degrees of freedom \cite{KT,HKT};
\item{} the gauge groups are restricted to $U(N)$ groups
\cite{Matsubara,Armoni};
\item{} fields can transform only in (anti-)fundamental, bi-fundamental
and adjoint representations\footnote{For example consider the rank-2 tensor
representation, $t^{ij}(x)$ of $U(N).$ The gauge transformation
would be $t^{ij}(x) \to U^{i}_{i'} * U^{j}_{j'} * t^{i' j'}.$
Because of the noncommutativity, this breaks the closure property,
$(t^U)^V=t^{U*V}.$}
of the gauge groups
\cite{Gracia-Bondia,Terashima,Chaichian:nogo};
\item{} the charges of matter fields are restricted to $0$ and $\pm 1$,
and this makes it difficult to give fractional electric charges to the quarks
\cite{Haya};
\item{} gauge anomalies cannot be cancelled in a chiral noncommutative
theory, hence the anomaly-free theory must be vector-like
\cite{Haya, Gracia-Bondia,Bonora:2000he}.
\end{enumerate}

The authors of Ref.~\cite{Chaichian:SM} made an important step in noncommutative
model building by
proposing a noncommutative model which satisfies criteria 2, 3 and 4.
Their model has the noncommutative gauge group
$U(3)\times U(2) \times U(1)$ with matter fields transforming
only in (bi-)fundamental representations, and remarkably, it predicts
the hypercharges of the Standard Model. In many respects their model
is similar to the bottom-up approach of \cite{bottomup}
to the string embedding of the Standard Model in purely commutative settings.
Unfortunately, the noncommutative $U(3)\times U(2) \times U(1)$ model
of \cite{Chaichian:SM} is affected by the UV/IR mixing which causes
the $U(1)$ hypercharge sector to decouple.

The motivation of this paper is to construct a noncommutative
embedding of the Standard Model which satisfies all the requirements listed above.
The model is based on the gauge group $U(4)\times U(3) \times U(2)$
with matter fields transforming
in noncommutatively allowed representations. In the infrared the gauge
group is spontaneously broken to the Standard Model group by a Higgs mechanism.
We need a larger gauge group than the authors of \cite{Chaichian:SM} in order
to incorporate the UV/IR mixing effects, yet remarkably we still find
the correct values of the hypercharges for all the fields of the Standard Model.

Noncommutative field theories are defined by replacing the ordinary
products of all fields in the Lagrangians of their commutative counterparts
by the star-products
\EQ{(\phi * \varphi) (x) \equiv \phi(x)\  e^{{i\over 2}\theta^{\mu\nu}
\stackrel{\leftarrow}{\partial_\mu}
\stackrel{\rightarrow}{\partial_\nu}} \  \varphi(x) \ . \label{stardef}}
In this way noncommutative theories can be viewed as field theories on
 ordinary commutative spacetime. For example,  the noncommutative pure
gauge theory action is
\EQ{
S_{\rm YM} [A] = -{1\over 2g^2}\int d^{4} x \ \Tr ( F_{\mu \nu}*  F_{\mu \nu}
 ) \ , \label{pureym}
}
where
the commutator in the field strength also contains the star-product.

The UV/IR mixing in noncommutative theories arises from the fact that
due to Eq.~\eqref{stardef}
certain Feynman diagrams acquire factors of the form
$e^{i k_\mu \theta^{\mu\nu} p_\nu}$
(where k is an external momentum and p is a loop momentum) compared to their commutative
counter-parts.\footnote{As it is customary in noncommutative literature, we
will call these diagrams nonplanar. At the same time, the diagrams in which all
the phase factors $e^{i k_\mu \theta^{\mu\nu} p_\nu}$ cancel, are called planar diagrams.}
At large values of the loop momentum $p$,
the oscillations of $e^{i k_\mu \theta^{\mu\nu} p_\nu}$
improve
the convergence of the loop integrals. However, as the external momentum vanishes, $k \to 0,$
the divergence reappears and
what would have been a UV divergence is now reinterpreted as an IR divergence instead.
This phenomenon of UV/IR mixing is specific to noncommutative theories and
does not occur in the commutative settings where the physics of high energy degrees
of freedom does not affect the physics at low energies.

Some of the earlier work on the
phenomenology of noncommutative theories
and on the noncommutative Standard  Model (e.g. \cite{CJSWW}) Taylor expands the exponential
vertex factors which misses the physics of the UV/IR mixing.
By Taylor-expanding
the star-products in the Lagrangian, one obtains the action
of the standard commutative theory plus
an infinite number of $\theta$-dependent higher-derivative
terms. At an energy-scale below the noncommutativity scale, $k^2\ll 1/\theta$,
the higher-derivative terms correspond to irrelevant operators.
One would naively expect that in the deep infrared one can simply drop
all the effects due to irrelevant operators.
This would imply that the noncommutative and the corresponding commutative
theories belong to the same universality class, i.e. in the infrared their
behaviour is identical. Classically the two theories are, in fact, identical
in this regime. But at quantum level, this universality is broken due to
the UV/IR mixing. The main point we are making, following \cite{KT,Minwalla,MST}
is that the effects of oscillating phases $e^{i k_\mu \theta^{\mu\nu} p_\nu}$
in Feynman diagrams can never be reproduced by the power-expansion in $\theta$.

Because of the breakdown of universality described above,
one cannot apply the conventional picture and simply decouple completely
the UV and IR sectors \cite{Minwalla}. However, one can still calculate the
Wilsonian effective action by integrating out the high-energy degrees of freedom
and keeping track of the UV/IR mixing effects -- following the approach initiated
in \cite{KT}.

The UV/IR mixing introduces new
infrared divergencies into certain sectors of gauge theories on noncommutative spaces.
In particular, this leads to quadratic and logarithmic IR divergencies in the
polarisation tensor of $U(1)$ gauge fields, \cite{MST} which
alter the dispersion relation of the photon. Fortunately, in a supersymmetric theory all
quadratic
divergences cancel and we are left only with logarithmic divergences \cite{MST,KT}.

The
remaining logarithmic divergencies were interpreted in \cite{KT,HKT} as new contributions to
the change of slope of the running coupling constant of the $U(1)$ sector
of $U(N)$ gauge theories.
In \cite{HKT} the low-energy Wilsonian effective
action for a large class of noncommutative supersymmetric theories was calculated
and the results showed that the UV/IR mixing occurs only for the U(1)
degrees of freedom which decouple (becoming unobservable) leaving a theory
which at low energies looks like a safe commutative SU(N) theory.
A similar decoupling between $U(1)$ and $SU(N)$ components of $U(N)$
was also observed
in \cite{Armoni} for the one-loop gluon propagator in
noncommutative QCD.

The conclusion one can draw from this is that it is conceivable to embed a
commutative $SU(N)$ theory, such as e.g. QCD or the weak sector of the Standard Model
into a supersymmetric noncommutative theory in the UV, but some extra care should be
taken with the QED $U(1)$ sector. It is, in fact, pretty clear that the UV/IR
mixing makes it impossible to interpret a noncommutative $U(1)$ theory as an
ultraviolet embedding of ordinary QED. The low-energy theory emerging from
the noncommutative $U(1)$ theory will become free in the 
extreme IR $k^2 \to 0$ (rather than just
weakly coupled) and in addition will have other pathologies.
When supersymmetric theories are softly-broken down to ${\cal N}=0$
non-logarithmic IR divergences can re-appear.
Models with the U(1) gauge group have been
analysed \cite{CCL,AGVM1,AGVM2} and tachyons can only be avoided if the model
has ${\cal N}=4$ supersymmetry; even in this case the tachyons
are avoided at the expense of giving a mass to the photon and fine tuning is
required to keep this below experimental limits. The prospects for
phenomenologically acceptable versions of such models looks bleak.

It is becoming pretty clear that the only realistic way to embed QED into
noncommutative settings is to recover the electromagnetic $U(1)$ from
a {\it traceless} diagonal generator of some higher $U(N)$ gauge theory.
The trace-$U(1)$ part of this theory will decouple in the IR due
to IR/UV mixing effects, and a traceless diagonal generator can give
$U(1)$ as well as some non-Abelian $U(n)$ factors in favourable settings.
So it seems that in order to embed QED into a noncommutative theory one should
learn how to embed the whole Standard Model.

In the following section we will show how the UV/IR mixing leads to
the decoupling of the overall $U(1)$ factors from the gauge groups
in the infrared. (It should be noted however, that this decoupling is 
logarithmic and hence, slow.)
In section 3 we will introduce the model, calculate the hypercharges,
and discuss the gauge-, the fermion- and the Higgs-sectors. We will also
outline how to cancel all the gauge anomalies by extending the model.

The model presented in this paper is one example of how the Standard Model
can be embedded into a microscopic noncommutative gauge theory. One particularly
interesting future direction would be to find a realistic supersymmetric
version which would exhibit a dynamical supersymmetry breaking.
This is motivated by the UV/IR-decoupled
$U(1)$ degrees of freedom which provide a natural candidate for the hidden sector
of dynamical supersymmetry breaking, as explained in \cite{CKT1}.

\section{UV/IR Mixing and the Decoupling of U(1)}

In this section we will briefly recall how the
UV/IR mixing effects in $U(N)$ noncommutative gauge theory
lead to a decoupling of the overall $U(1)$ factor
at energies below the noncommutative mass scale
$M_{\sst NC}\sim \theta^{-1/2}.$
Some more technical details related to
our treatment of the UV/IR mixing are
assembled in the Appendix which repeats the line of reasoning
initiated in \cite{KT,HKT}.

We will first consider an unbroken $U(N)$ noncommutative theory with all matter
fields transforming in the adjoint representation of the gauge group.
The UV/IR mixing effects are
present in the $U(1)$ sector,
but do not affect the $SU(N)$ degrees of freedom, such that
the
leading order terms in the derivative expansion of the Wilsonian
effective action read \cite{HKT}:
\EQ{
L_{\rm eff}  \ = \
-\frac{1  }{ 2g^{2}_{\sst U(1)}(k) }
 \  \Tr\, F^{\sst U(1)}_{\mu \nu}
F^{\sst U(1)}_{\mu \nu}
\ - \ \frac{1 }{ 2g^{2}_{\sst SU(N)}(k) }
 \ \Tr\,  F^{\sst SU(N)}_{\mu \nu}
F^{\sst SU(N)}_{\mu \nu}
\ + \
\cdots
\ ,
\label{uoneres}}
and the dots stand for terms involving matter fields and higher-derivative
corrections.
The multiplicative coefficients in front of the gauge kinetic terms in
\eqref{uoneres} define effective coupling constants of the corresponding
gauge factors at momentum scale $k$.
In the infrared we have effectively a matrix of $U(N)$ coupling constants:
\EQ{
\frac{1}{g_{\sst U(N)}^2 (k)}_{\sst[N^2]\times[N^2]} \ = \
\frac{1}{g^{ 2}_{\sst U(1)}(k)} \ \oplus \
\frac{1}{g_{\sst SU(N)}^2 (k)} \, \uno_{\sst[N^2-1]\times[N^2-1]} \
\label{wcfactwo}
}
The running of the
$SU(N)$ gauge coupling at 1-loop level is given by
the same standard expression as in the commutative case,
\EQ{
\frac{(4\pi)^2}{ g^{2}_{\sst SU(N)}(k)}  \, =\,
b_0 \, \log k^2  \ ,\label{rg3}}
where $b_0$ is the first coefficient of the beta-function of the $SU(N)$
gauge theory. At the same time, the running
of the $U(1)$ gauge coupling
has the asymptotic behaviour \cite{KT}:
\beqa
\frac{(4\pi)^2}{ g^{2}_{\sst U(1)}(k)}  &\rightarrow&
b_0\, \log k^2 \ , \qquad {\rm as} \
k^2\to\infty \ ,\label{rg1}
\\
\frac{(4\pi)^2}{ g^{2}_{\sst U(1)}(k)}
 &\rightarrow&
-b_0 \, \log { k^2} \ , \qquad {\rm as} \
k^2\to 0 \ ,
\label{rg2}
\eeqa
with the same $b_0$ as in \eqref{rg3}.

It follows from \eqref{rg3},\eqref{rg1},\eqref{rg2} that
the two effective coupling constants are identical in the UV
and run in opposite directions in the IR  as
a result of the UV/IR mixing affecting the $U(1)$ sector \cite{KT,HKT}.
This leads to a breakdown of noncommutative $U(N)$ gauge symmetry,
$U(N)\rightarrow U(1)\times SU(N),$ at momentum scales
$k\le M_{\sst NC}\sim \theta^{-1/2}.$ The $U(1)$ degrees of freedom
become weakly coupled and approach a free theory,
$g_{\sst U(1)} \rightarrow 0,$
as $k \rightarrow 0.$ The remaining $SU(N)$ degrees of freedom are
described, at energies below $M_{\sst NC},$ by the standard commutative gauge theory.
At the same time, in the UV region,
$k\gg M_{\sst NC},$ the full noncommutative $U(N)$ gauge
invariance is restored.

\begin{figure}
\centering
\includegraphics[width=10cm]{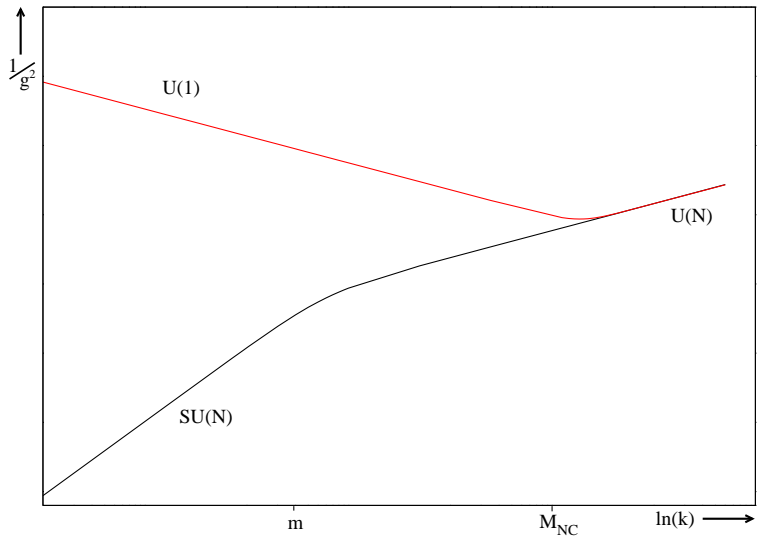}
\caption{\it The running of the couplings in a softly-broken \cal{N}=1 theory with
one chiral multiplet in the adjoint representation.}
\label{g_running}
\end{figure}

To illustrate these results,
Figure (\ref{g_running}) shows how the coupling varies in a softly-broken ${\cal N}=1$
supersymmetric gauge theory with one chiral multiplet of mass $m$ transforming
in the adjoint representation of $U(N)$.
As can be seen from the graph, in the infrared regime of the theory, the gauge boson
associated with the U(1) have decoupled \cite{HKT}.

It is worthwhile to note that what we have described here is
a dynamical breakdown of noncommutative $U(N)$ at low energies, which is induced
by different slopes of the running coupling constants.
It is expected that when the higher derivative
interactions are included in \eqref{uoneres},
the full noncommutative $U(N)$ gauge invariance will be recovered at the level of
the effective action.\footnote{This, however, does not affect our conclusions
about the different slopes in the running couplings and the decoupling of the $U(1)$.}
More precisely, effective actions
involving Wilson lines \cite{Das,Gross:observables,IIKK} can be
written down which are explicitly gauge invariant \cite{VanRaamsdonk:2001jd,AL,LT}
and which reduce to \eqref{uoneres} when higher-derivative terms are dropped.

We will now summarise the consequences of the UV/IR mixing for more general cases
relevant to considerations in this paper.
First, one can include $N_f$ flavours of matter fields transforming in the fundamental
representation\footnote{We recall that the only representations allowed in a noncommutative
gauge theory are adjoint, (anti)-fundamental and bi-fundamental ones. As we have already
included adjoint representations, and since bi- and anti-fundamental representations
are essentially the same as fundamental ones,
to cover the general case it is sufficient to add just fundamental representations.}
of the gauge group, and, second, the gauge group $U(N)$
can be spontaneously broken to $U(n)$ at a scale $m$, so that some
of the gauge bosons and matter fields become massive.

In the UV region, the theory is a noncommutative $U(N)$
and there is a single coupling constant,
\EQ{
\frac{(4\pi)^2}{ g^{2}_{\sst U(N)}(k)}  \, \rightarrow\,
b_0^{\sst U(N)\,N_f} \, \log k^2  \ ,
\qquad {\rm as} \
k^2\to\infty \ . \label{RGU}}
Here $b_0^{\sst U(N)\,N_f}$ is the 1-loop coefficient of the beta function
of the microscopic $U(N)$ theory with $N_f$ fundamental
flavours.\footnote{$b_0^{\sst U(N)\,N_f}$ takes the same
value as in the corresponding commutative $SU(N)$ theory.}
In the IR region, two things happen: the trace $U(1)$ factor decouples below
the noncommutative mass, and also, all the massive degrees of freedom freeze
at momentum scales below their masses,
\EQ{
\frac{1}{ g_{\sst U(N)}^2 (k)} \, \rightarrow\,
\frac{1}{g^{ 2}_{\sst U(1)}(k)} \ \oplus \
\frac{1}{g_{\sst SU(n)}^2 (k)} \, \uno_{\sst[n^2-1]\times[n^2-1]} \
\label{WCfactwo}
}
where
\beqa
\frac{(4\pi)^2}{ g^{2}_{\sst SU(n)}(k)}  &\rightarrow& \,
b_0^{\sst SU(n)}
\, \log k^2 \ , \label{RGSU}
\\
\frac{(4\pi)^2}{ g^{2}_{\sst U(1)}(k)}
 &\rightarrow& \,
-(2b_0^{\sst U(N)\,N_f=0}- b_0^{\sst U(N)\,N_f}
)\, \log { k^2} \ , \qquad {\rm as} \
k^2\to 0 \ .
\label{RGU1}
\eeqa
The UV/IR mixing affects only the $U(1)$ coupling and, hence,
the first equation \eqref{RGSU} takes the standard commutative
and recognizable form. However, the $U(1)$ coupling
is affected by the UV/IR mixing
and leads to the
slope in the IR given by $-2b_0^{\sst U(N)\,N_f=0} + b_0^{\sst U(N)\,N_f}$ as
follows from \eqref{RGU1}.
This expression for the slope follows the fact that the $\theta$-dependent
phase factors cancel in Feynman diagrams involving fundamental fields
propagating in the loop \cite{Haya} and do not cancel for adjoint fields in the loop.

Running couplings $\frac{1}{ g^2(k)}$ of noncommutative $U(1)$ (supersymmetric) theories
were first derived and plotted over the full range of the momentum scale $k$
in \cite{KT}.
Our expressions in \eqref{RGU},\eqref{RGU1} are in agreement with those results in the
asymptotic regions $k^2 \to \infty$ and $k^2 \to 0$. It should be noted that
expressions such as \eqref{RGU1} are valid in the extreme infrared,
at finite values of $k^2$ comparable to various mass scales in the theory,
the coupling changes slopes.

In a non-supersymmetric, or softly broken supersymmetric gauge theory there
are additional non-logarithmic sources of the UV/IR mixing effects
which modify the dispersion relation for (decoupled) the $U(1)$
gauge field. Following \cite{AGVM1}, it can be shown that these contributions
can be rendered harmless as long as two conditions are met.
Firstly, for each fermion
in the adjoint representation, there is a gauge field or complex scalar that is also in
the adjoint representation (and vice-versa). This condition removes quadratic
IR divergencies from the $U(1)$ polarisation tensor.
Secondly  the sum of the mass squared
for the adjoint fermions must be less or equal than the sum of the mass squared
for the complex scalars and the gauge bosons. If this condition is satisfied,
there are no tachyons in the decoupled $U(1)$ gauge sector.

In the model that we outline in the next section, we can satisfy both
these conditions hence the only consequence of the
UV/IR mixing is that the U(1) degrees of freedom decouple in the
extreme infrared. Even though the decoupling is only logarithmic,
in this paper we will always assume that these trace-$U(1)$ degrees of
freedom are completely decoupled and are 
essentially unobservable at
low-energies. In much of what will follow the overall $U(1)$ factors
of all three $U(N)$ gauge groups considered below will be dropped at
energies much below the noncommutative scale relevant to the commutative Standard Model.
The effects of these overall trace $U(1)$ degrees of freedom will be
discussed in future work \cite{new}.

\section{The Noncommutative Standard Model}

As was mentioned earlier, all fields in a Noncommutative Gauge Theory
must transform in the adjoint, fundamental, anti-fundamental or bi-fundamental
representations. We assign the fields to the representations shown in
table (\ref{reps_table}) and note that, unlike in the Standard Model, no field is
charged under more than two groups. All the matter fermion fields come in
three generations (which is not indicated explicitly in the table), and furthermore,
the table will be extended in section \ref{anomalies}.

\begin{table}[ht]
\centerline{
\begin{tabular}{|c||c|c|c|c|} \hline
Field      \xstrut   &   $U_{C}(4)$ & $U_{B}(3)$         & $U_{A}(2)$         & Hypercharge   \\ \hline
 $e_{R}$     \xstrut   &              &                    & $\Box$             & -2            \\ \hline
 $\nu_{R}$   \xstrut   &              &                    &                    &  0            \\ \hline
 $l_{L}=\left(\begin{matrix}\nu_{L} & e_{L}\end{matrix}\right)$
         \xstrut   &          &$\stackrel{-}{\Box}$&    $\Box$          & -1           \\ \hline
 $u_{R}$     \xstrut   &   $\Box$     &                &$\stackrel{-}{\Box}$& $+\frac{4}{3}$\\ \hline
 $d_{R}$     \xstrut   &   $\Box$     &                &                & $-\frac{2}{3}$\\ \hline
 $q_{L}=\left(\begin{matrix}u_{L} & d_{L}\end{matrix}\right)$
         \xstrut   &    $\Box$    &$\stackrel{-}{\Box}$&                    & $+\frac{1}{3}$\\ \hline
 $C_{\mu}$         \xstrut   &$\Box$$\stackrel{-}{\Box}$&        &                    &     0        \\ \hline
 $B_{\mu}$         \xstrut   &          &$\Box$$\stackrel{-}{\Box}$&              &     0        \\ \hline
 $A_{\mu}$         \xstrut   &          &                    &$\Box$$\stackrel{-}{\Box}$&     0          \\ \hline
 $\phi_{B}$  \xstrut   &          & $\Box$         &                    &     1        \\ \hline
 $\phi_{C\bar{B}}$
             \xstrut   &   $\Box$     &$\stackrel{-}{\Box}$&                    & $+\frac{1}{3}$\\ \hline
 $\phi_{B\bar{A}}$
             \xstrut   &              & $\Box$             &$\stackrel{-}{\Box}$& $1$\\ \hline
 $\phi_{A\bar{C}}$
                 \xstrut   &$\stackrel{-}{\Box}$&              &$\Box$ &  $-\frac{4}{3}$\\ \hline
         \end{tabular}}
 \caption{Representations for various fields in the theory}
 \label{reps_table}
 \end{table}

As can been seen from the table, we have introduced three
bi-fundamental Higgs fields, $\phi_{A\bar{C}}$,
$\phi_{C\bar{B}}$, $\phi_{B\bar{A}}$ and a fundamental $\phi_{B}$ compared
to the Standard Model's
single fundamental Higgs.
We note that unlike in \cite{Chaichian:SM}
the scalar Higgs fields in our model are proper fields defined
on the noncommutative space. Just like all other fields in the model,
Higgs fields appear in the Lagrangian with the star-products. This
is different from "Higgsac" fields used in \cite{Chaichian:SM}.
The latter have been
shown to violate unitary \cite{rizzo}.\footnote{These problems have recently
been addressed by utilizing semi-infinite Wilson lines in \cite{Chaichian:2004yw}.}

The scalar potential (discussed in section \ref{scalarpot}) will induce the
following VEV structure:
\begin{equation}
 \langle\phi_{C\bar{B}}\rangle= \left(
 \begin{matrix}
 0 & 0 & 0 \\
 0 & 0 & 0 \\
 0 & 0 & 0 \\
 0 & 0 & a
 \end{matrix} \right)
\quad
\langle\phi_{B\bar{A}}\rangle= \left(
 \begin{matrix}
 \tilde{v} & 0  \\
 0 & 0  \\
 0 & b  \\
 \end{matrix} \right)
\quad
\langle\phi_{A\bar{C}}\rangle= \left(
 \begin{matrix}
 0 & 0 & 0 & 0\\
 0 & 0 & 0 & c
 \end{matrix} \right)
\quad
\langle\phi_{B}\rangle= \left(
 \begin{matrix}
 0   \\
 v   \\
 0   \\
 \end{matrix} \right)
\label{vevs}
\end{equation}
The scalar potential will mean that the VEVs $a$, $b$ $c$ are much larger that
$v$ and $\tilde{v}$ which will turn out to be the electroweak breaking scale.

The gauge bosons for the groups $U_{C}(4)$, $U_{C}(3)$ and $U_{C}(2)$ are
respectively: $C^{p}_\mu$ ($p=0..15$), $B^{q}_\mu$ ($q=0..8$) and
$A^{r}_\mu$ ($r=0..3$).
So, for example, $\phi_{B}$ which transforms as
\begin{equation}
\phi_{B} \to U * \phi_{B} \phantom{with} U \in U_{B}(3)
\end{equation}
will have a covariant derivative:
\begin{equation}
D_{\mu} * \phi_{B} = \partial_{\mu} *\phi_{B} + i g_{B}B_{\mu}^{q}t^{q} * \phi_{B}
\end{equation}
and its Hermitian conjugate which transforms in the anti-fundamental, i.e.
\begin{equation}
\phi_{B}^{\dagger} \to \phi_{B}^{\dagger}*U^{-1} \phantom{with} U \in U_{B}(3)
\end{equation}
will have a covariant derivative:
\begin{equation}
D_{\mu} * \phi_{B}^{\dagger} = \partial_{\mu}* \phi_{B}^{\dagger} -
i g_{B}\phi_{B}^{\dagger} * B_{\mu}^{q}t^{q}
\end{equation}
In the following discussion we will neglect the generator of the trace $U(1)$
of each group ($C^{0}_\mu$, $B^{0}_\mu$ and $A^{0}_\mu$) for simplicity.
These generators
will decouple at low-energies from the effective commutative
Standard Model
due to UV/IR mixing.

It should be noted however that the decoupling of the trace $U(1)$ factors
from the $SU(N)$ degrees of freedom in the IR is logarithmic
$g^2_{\sst U(1)} \propto 1/\log k^2 \rightarrow 0,$ and hence the
$U(1)$ fields are decoupling slowly. It may well happen that
the effects of these extra $U(1)$ degrees of freedom are not negligible
and even important~\cite{new} for the low-energy physics at small non-vanishing
momenta. In the rest of this paper and in particular in
Section {\bf 3.1}, for simplicity
of presentation, we will always assume
that these $U(1)$ degrees of freedom have completely decoupled from the
low-energy effective theory.

\subsection{The Gauge Sector}

We start with the product of the three gauge groups,
$U_A(2) \times U_B(3) \times U_C(4).$
Their couplings are denoted as $g_{\sst A},$ $g_{\sst B},$  and $g_{\sst C},$
and their generators are $\hf \sigma^A,$ $\hf \lambda^B,$ and $\hf T^C$
respectively.
The vacuum expectation value for $\phi_{C\bar{B}}$ will partially break the
gauge group. The covariant derivative is
\begin{equation}
D_{\mu} \langle\phi_{C\bar{B}}\rangle = \partial_{\mu}\langle\phi_{C\bar{B}}\rangle
+ \frac{i}{2} g_{C} C_{\mu}^{p}T^{p} * \langle\phi_{C\bar{B}}\rangle
- \frac{i}{2} g_{B} \langle\phi_{C\bar{B}}\rangle * B_{\mu}^{q}\lambda^{q}
\end{equation}
where, as mentioned earlier we have neglected the trace-$U_C(1)$ and $U_B(1)$
fields $C_\mu^0$ and $B_\mu^0$ due to their decoupling at low energies caused
by the  UV/IR mixing. It is in principle straightforward to incorporate these trace-$U(1)$
fields in the analysis, but we will not pursue this in the present paper.
The $SU(4)$ generators $\hf T^{1\dots 15}$ are listed in
Appendix B and the $SU(3)$ generators $\hf \lambda^{1\ldots 8}$ are taken
to be the $\hf$ Gell-Mann matrices.

The $\left(D_{\mu} \langle\phi_{C\bar{B}}\rangle
\right)^{\dagger}\left(D_{\mu}\langle\phi_{C\bar{B}}\rangle\right)$
term in the Lagrangian will contain diagonal mass-terms:

\begin{equation}
\begin{aligned}
\frac{a^2}{4}\bigg(
g_{\sst C}^2\left((C_{\mu}^9)^2+(C_{\mu}^{10})^2+(C_{\mu}^{11})^2+(C_{\mu}^{11})^2+(C_{\mu}^{13})^2+(C_{\mu}^{14})^2 \right) \\
+g_{\sst B}^2\left((B_{\mu}^4)^2+(B_{\mu}^{5})^2+(B_{\mu}^{6})^2+(B_{\mu}^{7})^2 \right)
\bigg)
\end{aligned}
\end{equation}

and non-diagonal mass-terms:
\begin{equation}
\frac{a^2}{4}\left(
\frac{3}{2}g_{\sst C}^2(C_{\mu}^{15})^2+\frac{4}{3}g_{\sst B}^2(B_{\mu}^8)^2- 2
\sqrt(2)g_{\sst B} g_{\sst C} B_{\mu}^8 C_{\mu}^{15}
\right)
\end{equation}
If we rotate to a new basis:
\begin{equation}
\left(\begin{matrix}
M_{\mu}^1 \\
M_{\mu}^2
\end{matrix}\right)
=
\left(\begin{matrix}
\mbox{cos} \, \theta_{CB}  & \mbox{sin}\,  \theta_{CB}\\
-\mbox{sin}\,  \theta_{CB} & \mbox{cos} \, \theta_{CB}
\end{matrix}\right)
\left(\begin{matrix}
C_{\mu}^{15} \\
B_{\mu}^8
\end{matrix}\right)
\end{equation}
where
\begin{equation}
\mbox{cos} \, \theta_{CB} = \frac{ \sqrt{2} g_{\sst B}}{\sqrt{2 g_{\sst B}^2 + \frac{9}{4} g_{\sst C}^2}}
\qquad
\mbox{sin} \, \theta_{CB} = \frac{ \frac{3}{2} g_{\sst C}}{\sqrt{2 g_{\sst B}^2 +
\frac{9}{4} g_{\sst C}^2}}
\label{CB_angle}
\end{equation}
Then $M_{\mu}^1$ will be massless but $M_{\mu}^2$ will acquire
a mass so, out of the $U(4)\times U(3)$ that we start with, the following
gauge bosons are still massless: $C^{1..8}_{\mu}$ (which we will identify with
the $SU(3)_C$ of the Standard Model, $B^{1..3}_{\mu}$ (which we will identify
with $SU(2)_L$) and $M^{1}_{\mu}$.

The covariant derivative for $\phi_{B\bar{A}}$ will lead to a term involving its
vacuum expectation value. Because $\tilde{v}<<b$ (in equation \eqref{vevs}) we
will temporarily set $\tilde{v} \to 0$
\begin{equation}
D_{\mu} \langle\phi_{B\bar{A}}\rangle = \partial_{\mu} \langle\phi_{B\bar{A}}\rangle
+ \frac{i}{2} g_{B} B^{q}_\mu\lambda^{q} * \langle\phi_{B\bar{A}}\rangle
- \frac{i}{2} g_{A} \langle\phi_{C\bar{B}}\rangle * A^{r}_\mu \sigma^{r}
\end{equation}
where the $SU(2)$ generators $\hf \sigma^r$ are the usual
$\hf$ Pauli matrices.
However, $B^8_{\mu}=\mbox{sin} \, \theta_{CB} M_{\mu}^1 +\mbox{cos} \, \theta_{CB} M_{\mu}^2$
so ignoring the massive gauge bosons we have:
\begin{equation}
D_{\mu} \langle\phi_{B\bar{A}}\rangle = \partial_{\mu} \langle\phi_{B\bar{A}}\rangle
+ \frac{i}{2} g_{B} B_{\mu}^{q}\lambda^{q} * \langle\phi_{B\bar{A}}\rangle
+ \frac{i}{2} g_{0} M_{\mu}^{1}\lambda^{8} * \langle\phi_{B\bar{A}}\rangle
- \frac{i}{2} g_{A} \langle\phi_{C\bar{B}}\rangle * A_{\mu}^{r}\sigma^{r}
\end{equation}

where $g_{0}=g_{B}\mbox{sin} \, \theta_{CB}$ and now $q,r=1..3$.

The resulting diagonal mass-terms will be:
\begin{equation}
\frac{b^2}{4}\left(
g_{\sst A}^2\left((A_{\mu}^1)^2+(A_{\mu}^{2})^2\right)
\right)
\end{equation}
and the remaining mass-terms are:

\begin{equation}
\frac{b^2}{4}\left(
g_{\sst A}^2(A_{\mu}^3)^2+\frac{4}{3}g_0^2(M_{\mu}^{1})^2 - \frac{4}{\sqrt{3}}g_0 g_{\sst A}(A_{\mu}^3)(M_{\mu}^{1})
\right)
\end{equation}

We can diagonalise these by writing:
\begin{equation}
\left(\begin{matrix}
Y_{\mu}\\
M_{\mu}^3
\end{matrix}\right)
=
\left(\begin{matrix}
\mbox{cos} \, \theta_{BA}  & \mbox{sin}\,  \theta_{BA}\\
-\mbox{sin}\,  \theta_{BA} & \mbox{cos} \, \theta_{BA}
\end{matrix}\right)
\left(\begin{matrix}
A_{\mu}^{3} \\
M_{\mu}^1
\end{matrix}\right)
\end{equation}
where
\begin{equation}
\mbox{cos} \, \theta_{BA} = \frac{ \sqrt{3} g_{\sst A}}{\sqrt{3 g_{\sst A}^2 + 4 g_0^2}}
\qquad
\mbox{sin} \, \theta_{BA} = \frac{ \frac{3}{2} g_{\sst C}}{\sqrt{3 g_{\sst A}^2 + 4 g_0^2}}
\label{BA_angle}
\end{equation}

The field labelled $Y_{\mu}$ is the gauge boson for a massless $U(1)$ and
will be identified with the hypercharge whilst the $M_{\mu}^3$ field has acquired a mass.

If we now calculate which gauge degrees of freedom are given a mass by
$\phi_{A\bar{C}}$ it will turn out that no massless degrees of freedom
acquire a mass; the gauge group is broken no further. In particular the
$Y_{\mu}$ field remains unchanged.

To summarise, after decoupling the trace-$U(1)$ factors
the group $SU_C(4)$ has been broken to $SU(3) \times U(1)$, $SU_B(3)$ has
been broken to $SU(2)_L\times U(1)$ and $U_A(2)$ has been broken to $U(1)$. All of these
$U(1)$ factors arise from traceless generators of $SU$ groups.
Furthermore, only one linear
combination of these three traceless $U(1)$'s remains massless. This single $U(1)$ group
will now be identified with the hypercharge of the Standard Model.

\subsection{Hypercharges}

The hypercharge for each particle is determined by the representation of the particle under the
microscopic gauge groups. The ideas in this section follow \cite{bottomup,Chaichian:SM} but because
of our unusual gauge-group ($U(4) \times U(3) \times U(2)$) the details differ.

The coupling of the right handed electron (c.f table \ref{reps_table}) to the hypercharge is determined by:
\begin{align}
\bar{e}_R \gamma^{\mu} D_{\mu} e_R =& \bar{e}_R \gamma^{\mu}\partial_{\mu} e_R
+\frac{i}{2}g_{\sst A} \bar{e}_R \gamma^{\mu} e_R A^3_{\mu} \sigma^3 + \frac{i}{2}g_{\sst A}
\bar{e}_R \gamma^{\mu} [e_R, A^3_{\mu}] \sigma^3\\
&\bar{e}_R \gamma^{\mu}\partial_{\mu} e_R
+\frac{i}{2}g_{\sst A} \bar{e}_R \gamma^{\mu} e_R \mbox{sin} \, \theta_{BA} Y_{\mu} \sigma^3
\end{align}

We have ignored the $[e_R, A^3_{\mu}]$ term as we are considering scales well below the noncommutative scale. The coupling between
$Y_{\mu}$ and the particle in the first row of the U(2) doublet is therefore $g_{\sst A} \mbox{sin} \, \theta_{BA}$. This should be
proportional to the hypercharge, $-2g'$ where we define $g' \equiv \frac{1}{2} g_{\sst A} \mbox{sin}  \, \theta_{BA}$ to be the
coupling to the hypercharge. With this definition, the hypercharge of all the other particles in the model is now fixed.

The right-handed down quark transforms in the fundamental of $U_C(4)$:

\begin{equation}
\bar{d}_R \gamma_{\mu} D_{\mu} d_R = \bar{d}_R \gamma_{\mu} \partial_{\mu} d_R
+\frac{i}{2}g_{\sst C}\bar{d}_R \gamma_{\mu} d_R T_{15} C_{\mu}^{15}+...
\end{equation}

Writing $C_{\mu}^{15} = \mbox{cos}\, \theta_{CB} M_{\mu}^1 - \mbox{sin} \,\theta_{CB} M_{\mu}^2$ and then
$M_{\mu}^1 = \mbox{cos} \, \theta_{BA} Y_{\mu} - \mbox{sin}\, \theta_{BA} N_{\mu}^3$ then the term that will
determine the coupling is:

\begin{equation}
\frac{i}{2}g_{\sst C} \bar{d}_R \gamma^{\mu} d_R T_{15}\mbox{cos}\, \theta_{CB} \mbox{cos} \, \theta_{BA} Y_{\mu}
\end{equation}

So the coupling for the right-handed down quark (for all except the fourth particle in the multiplet) is:
\begin{equation}
Y_{d_R} = \frac{g_{\sst C}}{\sqrt{6}} \mbox{cos}\, \theta_{CB} \mbox{cos} \, \theta_{BA}
\end{equation}
Using equations \eqref{CB_angle} and \eqref{BA_angle} we find $\frac{Y_{d_R}}{g'}=\frac{-2}{3}$ i.e. the
Standard Model value.

We can calculate the hypercharges of the other particles in an analogous fashion. For example the
multiplet of left-handed leptons has a term which can be written (ignoring massive fields):

\begin{align}
&=\frac{i}{2}\bar{\psi}_L^l \gamma_{\mu} \psi_L^l (g_{\sst A} \sigma^3 A_{\mu}^3 - g_{sst B} B_{\mu}^{8} \lambda^8) \\
&=\frac{i}{2}\bar{\psi}_L^l \gamma_{\mu} \psi_L^l (g_{\sst A} \sigma^3 \mbox{sin} \, \theta_{BA} Y_{\mu}
- g_{sst B} \lambda^8 \mbox{sin} \, \theta_{CB} M_{\mu}^1) \nonumber \\
&=\frac{i}{2}\bar{\psi}_L^l \gamma_{\mu} \psi_L^l (g_{\sst A} \sigma^3 \mbox{tan} \, \theta_{BA} \mbox{cos} \, \theta_{BA}
- g_{sst B} \lambda^8 \mbox{sin} \, \theta_{CB} \mbox{cos} \, \theta_{BA}) Y_{\mu} \nonumber
\end{align}

So the hypercharge of the left-handed leptons will be:
\begin{align}
\frac{Y_{\psi_L^l}}{g'} &=
\frac{\left(g_{\sst A} \mbox{tan} \, \theta_{BA} - \frac{g_{sst B}}{\sqrt{3}} \mbox{sin} \, \theta_{CB}\right)\mbox{cos} \, \theta_{BA}}
{\frac{1}{2}g_{\sst A} \mbox{cos} \, \theta_{BA} \mbox{tan} \, \theta_{BA}} \\
&=-1 \nonumber
\end{align}

The hypercharges of the other fields is listed in table (\ref{reps_table}) and each agrees
with its Standard Model value.

\subsection{Anomalies and Extra Fields}
\label{anomalies}

Anomalies have been thoroughly studied in a noncommutative context
\cite{Haya, Gracia-Bondia,Bonora:2000he,Intriligator:2001yu}.
The generally accepted conclusion is that in order for a noncommutative theory to be free
of chiral anomalies, the theory must be vector-like.

The matter content introduced so far is chiral, as it must be in order to match the
Standard Model matter content at low energies; we have left-handed (but no right-handed)
fermions under the $U_B(3)$ gauge group that will become the $SU(2)_L$ group in the
low-energy limit of the theory. To fix the problem we introduce three extra heavy
generations, one for each observed generation. Each particle in these heavy generations
must have the opposite chirality to their Standard Model counterpart.

Although these extra generations circumvent the problems with anomalies, this might also
be possible by adding fewer fields to the theory. However, in the next section we will
see that these extra heavy generations are essential when writing down the necessary
Yukawa terms for the theory.

We must also add three more fields to the model. As discussed earlier there are two conditions
that need to prevent quadratic divergences arising in the polarisation tensor of the decoupled
U(1) gauge bosons. The first condition is that for each fermion in the adjoint representation
of a gauge group, there is a gauge field or complex scalar that is also in the adjoint
representation (and vice-versa). We have adjoint gauge fields but no adjoint matter
so we add
massive adjoint fermion fields, one per microscopic gauge group;
 $\lambda^{A}$, $\lambda^{B}$ and
$\lambda^{C}$.
Table \ref{extra_reps} summarises the extra matter we have needed to add to the
theory.

\begin{table}[ht]
\centerline{
\begin{tabular}{|c||c|c|c|c|} \hline
Field      \xstrut   &   $U_{C}(4)$ & $U_{B}(3)$         & $U_{A}(2)$         & Hypercharge   \\ \hline
 $E_{L}$     \xstrut   &              &                    & $\Box$             & -2            \\ \hline
 $N_{L}$   \xstrut   &              &                    &                    &  0            \\ \hline
 $L_{R}=\left(\begin{matrix}N_{R} & E_{R}\end{matrix}\right)$
         \xstrut   &          &$\stackrel{-}{\Box}$&    $\Box$          & -1           \\ \hline
 $U_{L}$     \xstrut   &   $\Box$     &                &$\stackrel{-}{\Box}$& $+\frac{4}{3}$\\ \hline
 $D_{L}$     \xstrut   &   $\Box$     &                &                & $-\frac{2}{3}$\\ \hline
 $Q_{R}=\left(\begin{matrix}U_{R} & D_{R}\end{matrix}\right)$
         \xstrut   &    $\Box$    &$\stackrel{-}{\Box}$&                    & $+\frac{1}{3}$\\ \hline
 $\lambda^{C}$         \xstrut   &$\Box$$\stackrel{-}{\Box}$&        &                    &     0        \\ \hline
 $\lambda^{B}$         \xstrut   &          &$\Box$$\stackrel{-}{\Box}$&              &     0        \\ \hline
 $\lambda^{A}$         \xstrut   &          &                    &$\Box$$\stackrel{-}{\Box}$&     0          \\ \hline
\end{tabular}}
 \caption{Representations for the "extra" fields added to the theory}
 \label{extra_reps}
 \end{table}

\subsection{Yukawa Couplings}

Unlike the Standard Model, multiple Higgs fields are required in order
give mass to all the particles. The Yukawa terms can be arranged into two
categories. Firstly, there are terms that involve fields from the same generation.
Secondly, because we have generations with opposite chirality, we have novel terms
involving fields from different generations. Additionally, as in the Standard Model,
there can be the usual mixing between the generations but we neglect these here for
simplicity.

Yukawa terms of the first type are (for one light generation):
\begin{equation}
\nu_R^{\dagger} \phi_{B \bar{A}} l^L_{A\bar{B}}+ e^R_{A} \phi^{\dagger}_{\bar{B}} {l_L^{\dagger}}_{B \bar{A}}
+ q^L_{C \bar{B}} \phi_{B \bar{A}} {u_R^{\dagger}}_{A \bar{C}} + d^R_{C} \phi^{\dagger}_{\bar{B}} {q_L^{\dagger}}_{B \bar{C}}
+ \mbox{h.c}
\end{equation}
and for a heavy generation:
\begin{equation}
N_L^{\dagger} \phi_{B \bar{A}} L^R_{A\bar{B}}+ E^L_{A} \phi^{\dagger}_{\bar{B}} {L_R^{\dagger}}_{B \bar{A}}
+ Q^R_{C \bar{B}} \phi_{B \bar{A}} {U_L^{\dagger}}_{A \bar{C}} + D^L_{C} \phi^{\dagger}_{\bar{B}} {Q_R^{\dagger}}_{B \bar{C}}
+ \mbox{h.c}
\end{equation}

These terms on their own are not sufficient to give large masses to all those
particles which are not observed at low energies, for example there is a fourth
"colour" of quark that would not interact with the strong force, as the gauge
group has been broken from $SU(4)$ to $SU(3)$ but would still interact
electromagnetically.

The extra three generations in which each particle has the opposite chirality to
its Standard Model equivalent (as introduced in section \ref{anomalies} to
cure the problems with anomalies) also cures the problem here. The possible terms that
mix a light generation with a heavy generation are:

\begin{equation}
{E_L^{\dagger}}_{\bar{A}}\phi_{A \bar{C}}d^{R}_C+{L_{R}^{\dagger}}_{B \bar{A}} \phi_{A \bar{C}} q^{L}_{C \bar{B}}
+N_L^{\dagger} \phi_{A \bar{C}}  u^{R}_{C\bar{A}} + l^{L}_{A \bar{B}} \phi^{\dagger}_{B \bar{C}} {U_{L}^{\dagger}}_{C\bar{A}}
+ q^{L}_{C \bar{B}} \phi^{\dagger}_{B \bar{C}} N^{L} + \mbox{h.c.}
\end{equation}

Notice that in the above generation-mixing terms (which violate baryon and lepton number)
neither of the Higgses with an electroweak scale vacuum expectation appear, so leptoquark
would only occur at a high energy scale, characterised by the $a$, $b$ and $c$ vacuum
expectation values.

When all possible such Yukawa terms are included, the particle content of the model at low
energies agrees with the observed spectrum of particles. Moreover the form of the coupling
gives a natural explanation for the extremely small mass of the left-handed neutrino
in the three light generations, the see-saw effect will naturally suppress their masses
to be of order $\tilde{v}^2/a$ although there are enough parameters to keep the neutrinos
in the three extra generations above the experimental bounds.

\subsection{The Higgs Potential}
\label{scalarpot}

The pattern of symmetry breaking and mass splittings in the preceding sections
was dependent on a particular pattern \eqref{vevs}
of vacuum expectation values for the Higgs fields.
We now will construct a simple example of the scalar potential which generates
the vev structure in Eq.~\eqref{vevs}.

First, using gauge transformations $SU_B(3)$,
we put $\phi_B$ in the canonical form:
\EQ{
\langle\phi_{B}\rangle= \left(
 \begin{matrix}
 0  \\
 v  \\
 0   \\
 \end{matrix} \right)
\ . \label{phib}
}
Next, we require that $\phi^{\dagger}_{\bar{B}} \phi_{B \bar{A}}=0$ and
further use the $SU_A(2)$ and $SU_B(2)$ transformations\footnote{$SU_B(2)$
is the subgroup of $SU_B(3)$ which leaves \eqref{phib} invariant.}
to diagonalise
$\phi_{B\bar{A}}$:
\EQ{
\langle\phi_{B\bar{A}}\rangle= \left(
 \begin{matrix}
 \tilde{v} & 0  \\
 0 & 0  \\
 0 & b  \\
 \end{matrix} \right)
 \ . \label{phiba}
}
Next we turn to $\phi_{C\bar{B}}$ and require that
$\phi_{C \bar{B}} \phi_{B}=0.$ We also use $SU_C(4)$ to simplify
$\phi_{C\bar{B}}$ further, such that:
\EQ{
\langle\phi_{C\bar{B}}\rangle= \left(
 \begin{matrix}
 0 & 0 & 0 \\
 0 & 0 & 0 \\
 0 & 0 & f \\
 g & 0 & a
 \end{matrix} \right)
 \ . \label{phicb}
}
At this stage to achieve relatively simple expressions in
\eqref{phib},\eqref{phiba},\eqref{phicb},
we have used all the available gauge symmetry
and the orthogonality conditions which
follow from the potential:
\begin{equation}
\lambda_1|\phi^{\dagger}_{\bar{B}} \phi_{B \bar{A}}|^2 +
\lambda_2|\phi_{C \bar{B}} \phi_{B}|^2 \ .
\end{equation}
This essentially leaves the third bi-fundamental Higgs
unrestricted at this stage,
\EQ{
\langle\phi_{A\bar{C}}\rangle= \left(
 \begin{matrix}
 z_1   & z_2    & z_3   & z_{4}\\
 z_{5} & z_{6} & z_{7} & c
 \end{matrix} \right) \ . \label{phiac}
}
Before imposing restrictions on $\phi_{A\bar{C}}$, we would like to
first further simplify the expressions \eqref{phib},\eqref{phiba},\eqref{phicb}.

We introduce another term in the scalar potential,
\begin{equation}
\lambda_3 \left({\cal D}_{B \bar{B}} - \mu_1^2 \uno_{B \bar{B}} \right)^2
\ , \label{Dterm}
\end{equation}
where
${\cal D}_{B \bar{B}}$ is a bilinear combination of Higgs fields:
\begin{equation}
{\cal D}_{B \bar{B}} \equiv \phi_{B \bar{A}}
\phi^{\dagger}_{A \bar{B}} - \phi^{\dagger}_{B \bar{C}}
\phi_{C \bar{B}} + \phi_B \phi^{\dagger}_{\bar{B}}
\ . \label{Dfield}
\end{equation}
On the right hand side of \eqref{Dfield} the indices $\bar{A},A$ and
$\bar{C},C$ are summed over, but not the indices $\bar{B},B$ which are
left free, so that ${\cal D}_{B \bar{B}}$
transforms in the adjoint of $SU_B(3)$. The scalar potential \eqref{Dterm}
contains a trace over gauge indices, hence
$\bar{B},B$  are finally summed over, and the Higgs potential
\eqref{Dterm} is a gauge singlet.

At the minimum of the potential \eqref{Dterm} we have,
\begin{equation}
\left(\begin{matrix}
|\tilde{v}|^2-|g|^2 & 0  & -g^{\dagger} a \\
0 & |v|^2 & 0 \\
-g a^{\dagger} & 0 & |b|^2 - |a|^2 - |f|^2
\end{matrix}\right)
=\left(\begin{matrix}
\mu_1^2 & 0 & 0 \\
0 & \mu_1^2 & 0 \\
0 & 0 & \mu_1^2
\end{matrix} \right) \ ,
\end{equation}
and the vacuum solution is:
\EQ{
g=0 \ , \quad |\tilde{v}|^2=\mu_1^2=|v|^2 \ , \quad
|b|^2 - |a|^2 - |f|^2 = \mu_1^2 \ .
\label{solone}
}

We continue reducing the number of free parameters in the vev
structure in a similar way to the considerations above and introduce
another term in the scalar potential:
\begin{equation}
\lambda_4 \left({\cal E}_{C \bar{C}} - \mu_2^2 \uno_{C \bar{C}} \right)^2 \ ,
\label{pot4}
\end{equation}
where ${\cal E}_{C \bar{C}}$ is defined as
\begin{equation}
{\cal E}_{C \bar{C}} \ \equiv \  \phi_{C \bar{B}}\phi^{\dagger}_{B \bar{C}} \ =\
\left(\begin{matrix}
0 & 0 & 0 & 0 \\
0 & 0 & 0 & 0 \\
0 & 0 & |f|^2 & f a^{\dagger} \\
0 & 0 & f^{\dagger} a & |a|^2
\end{matrix}\right) \ .
\end{equation}
The potential \eqref{pot4} is minimal at
\EQ{
a=\mu_2 \ , \qquad f=0 \ ,
\label{soltwo}
}
which complements the configuration \eqref{solone}.

We now return to the so far unconstrained Higgs field \eqref{phiac}
and write down new terms in the scalar potential
\begin{equation}
\lambda_5 \left({\cal G}_{A \bar{A}} - \mu_1^2 \uno_{A \bar{A}} \right)^2 \ + \
\lambda_6 \left({\cal K}_{C \bar{C}} \right)^2 \ ,
\label{pot56}
\end{equation}
where
\EQ{
{\cal G}_{A \bar{A}} \equiv
- \phi_{A \bar{C}}\phi^{\dagger}_{C \bar{A}}
+ \phi^{\dagger}_{A \bar{B}} \phi_{B \bar{A}}
\ , \quad
{\cal K}_{C \bar{C}} \equiv  \phi_{C \bar{B}}\phi^{\dagger}_{B \bar{C}}
- \phi^{\dagger}_{C \bar{A}} \phi_{A \bar{C}} \ .
}
The minimum of \eqref{pot56} is
\EQ{
z_1=z_2=z_3=z_4=z_5=z_6=z_7=0 \ , \quad |c|^2=|a|^2=\mu_2^2 \ , \quad f=0 \ .
}

The combined vacuum configuration gives
\begin{equation}
 \langle\phi_{C\bar{B}}\rangle= \left(
 \begin{matrix}
 0 & 0 & 0 \\
 0 & 0 & 0 \\
 0 & 0 & 0 \\
 0 & 0 & \mu_2
 \end{matrix} \right)
\quad
\langle\phi_{B\bar{A}}\rangle= \left(
 \begin{matrix}
 \mu_1 & 0  \\
 0 & 0  \\
 0 & \sqrt{\mu_2^2+\mu_1^2}  \\
 \end{matrix} \right)
\quad
\langle\phi_{A\bar{C}}\rangle= \left(
 \begin{matrix}
 0 & 0 & 0 & 0\\
 0 & 0 & 0 & \mu_2
 \end{matrix} \right)
\quad
\langle\phi_{B}\rangle= \left(
 \begin{matrix}
 0   \\
 \mu_1   \\
 0   \\
 \end{matrix} \right)
\nonumber
\end{equation}
which reproduces \eqref{vevs}.

\vskip 1truecm

\centerline{\Large \bf Acknowledgements}

We are grateful to
Steve Abel, Anthony Owen, Andreas Ringwald, and especially Gabriele Travaglini
for useful discussions. JL was supported by a PPARC Studentship and VVK
by a PPARC Senior Fellowship.

\newpage

\startappendix
\Appendix{UV/IR and the Polarisation Tensor}

Our discussion here follows closely
the formalism introduced in
\cite{KT,HKT}, to which we refer the reader for further details.
To derive the matrix of effective coupling constants, Eq.~\eqref{wcfactwo},
we use a background perturbation theory and decompose
the gauge field $A_\mu$ into a background field
$B_\mu$ and a fluctuating quantum field $N_\mu$,
\EQ{
\label{fdec}
A_\mu= B_\mu + N_\mu
\ \ .
}
The
effective action $S_{\rm eff}(B)$  is obtained by functionally
integrating over the fluctuating fields.
Gauge invariance
constrains the interactions which can be generated
in this procedure. Therefore, the
effective action will always contain the kinetic term
\EQ{
S_{\rm eff} ( B ) \ni - \frac{1}{2g^2_{\rm eff}}
\int d^{4} x \ \Tr \left( F_{\mu \nu} *  F_{\mu \nu} \right)
\ \ .
}
The factor $\frac{1}{2g^2_{\rm eff}}$ on the right hand side is identified
with the effective coupling constant at the momentum scale $k$ of
the background field $B_\mu$.
In order to determine $g_{\rm eff}$ it is sufficient
to consider  the kinetic term $(\partial_\mu B_\nu - \partial_\nu B_\mu)^2$.
In the effective Lagrangian, this term
becomes
\EQ{
2 \int \frac{d^4 k}{(2\pi)^4} B_\mu^{A} (k) B_\nu^{B} (-k)
\ \Pi_{\mu \nu}^{AB} (k)
\ \ .
\label{wptdef}
}
Equation (\ref{wptdef}) defines the polarization tensor $\Pi_{\mu \nu}^{AB} (k)$,
which in the effective theory replaces the tree level transverse tensor
$k^2 \delta_{\mu\nu}-k_\mu k_\nu $.
On general grounds, $\Pi_{\mu \nu}^{AB} (k)$ has the structure
\EQ{
\Pi_{\mu \nu}^{AB} (k) =
\Pi_1^{AB} (k^2, \tilde{k}^2) (k^2 \delta_{\mu\nu}-k_\mu k_\nu ) +
\Pi_2^{AB} (k^2, \tilde{k}^2) \frac{\tilde{k}_\mu\tilde{k}_\nu }{ \tilde{k}^4}
\ \ .
\label{pimunu}
}
The matrix of the running coupling constants in \eqref{wcfactwo}
is determined entirely by
$\Pi_1^{AB}(k^2, \tilde{k}^2)$  via
\EQ{
\left[\frac{1}{ g_{\rm eff}^2 (k)}\right]^{AB} \ = \
\frac{\delta^{AB}}{ g_{\rm micro}^2 } \ + \
4 \Pi_1^{AB} (k^2, \tilde{k}^2) \ . \label{A.5}
}

The term in \eqref{pimunu}
proportional to $\tilde{k}_\mu\tilde{k}_\nu /\tilde{k}^4 $
would not appear in
ordinary commutative theories. It is transverse and
has derivative
dimension $-2$; therefore it is of leading order
compared to the standard gauge-kinetic term
(which has  derivative dimension $+2$), and
leads to a power-like infrared singular behaviour.
It is known
that $\Pi_2$ vanishes
for supersymmetric noncommutative gauge theories,
 as was first discussed
in \cite{MST}. For nonsupersymmetric theories,
$\Pi_2$ can potentially present serious problems.
For our purposes however, it will be sufficient to note that for noncommutative theories
with a matching number of bosonic and fermionic degrees of freedom
transforming in the adjoint representation of $U(N)$, the term
$\Pi_2$ is rendered harmless if a certain mass inequality relation is
satisfied, \cite{AGVM1}. Hence, for the rest of this Appendix we will concentrate
mostly on $\Pi_1$.

The action functional which describes the dynamics of
a spin-$j$ noncommutative field in the representation
{\bf r} of the gauge group in the background of $B_\mu$ has the general
form \cite{KT,Peskin}
\beqa
S [ \phi ] &=&
-\int d^{4} x \  \phi_{m,a} *
\left ( - D^2 (B)\delta_{mn}\delta^{ab} + 2i (F_{\mu \nu}^B)^{ab}
\hf J^{\mu \nu}_{mn}
\right) * \phi_{n,b}
\cr
&\equiv&-\int d^{4} x \
\phi_{m,a} * [ \Delta_{j,{\bf r}}]_{mn}^{ab} * \phi_{n,b}
\ \ .
\eeqa
Here $a,b$ are indices of the  representation {\bf r} of noncommutative $U(N)$,
$F^{ab}\equiv \sum_{A=1}^{N^2} F^A t^A_{ab}$, and $m,n$ are spin indices and
$J^{\mu \nu}_{mn}$ are
the generators of the euclidean Lorentz group appropriate
for the spin of $\phi$:
\beqa
J &=& 0 \qquad \qquad \qquad \qquad {\rm for\ spin\ 0 \ fields},
\\ \nonumber
J^{\mu \nu}_{\rho \sigma}&=&i(\delta^\mu_\rho \delta^\nu_\sigma -
\delta^\nu_\rho\delta^\mu_\sigma) \qquad {\rm for\ 4\hbox{-}vectors},
\\ \nonumber
[J^{\mu \nu}]_{\alpha}^{\ \beta} &=&
i  \hf [\sigma^{\mu \nu}]_{\alpha}^{\ \beta}
\qquad \qquad {\rm for\ Weyl\ fermions}\ \ .
\eeqa

At the one-loop level, the effective action is given by
\EQ{
\label{noncera}
S_{\rm eff} [B] =
-\frac{1  }{ 2g^2}
\int d^{4} x \ \Tr F_{\mu \nu} *  F_{\mu \nu} -
\sum_{j,{\bf r}} \alpha_{j} \log {\rm det}_* \Delta_{j, {\bf r}}
\ \ ,
}
where the sum is extended to all fields in the theory, including ghosts and gauge fields.
\begin{table}[ht]
\centerline{
\begin{tabular}{|c||c|c|c|c|} \hline
  \xstrut   & ghost & real scalar & Weyl fermion & gauge field   \\ \hline
 $\alpha_j$ \xstrut & 1&$-\frac{1}{2}$&$\frac{1}{2}$&$-\frac{1}{2}$ \\ \hline
 $d(j)$  \xstrut   & 1& 1        &    2    &     4    \\ \hline
 $C(j)$  \xstrut   & 0& 0        &$\frac{1}{2}$&     2    \\ \hline
 \end{tabular}}
 \caption{Constants for the various fields in the theory}
 \label{const_table}
 \end{table}

Functional star-determinants are computed by
\SP{\log {\rm det}_{*} \Delta_{j, {\bf r}} \equiv&
\log {\rm det}_{*} (-\partial^2 + {\cal K}(B)_{j, {\bf r}})\\
&= \log {\rm det}_{*} (-\partial^2) +
{\rm tr}_{*} \log (1+(-\partial^2)^{-1}{\cal K}(B)_{j, {\bf r}}) \ .
\label{logkdef}
}
The first term on the second line of \eqref{logkdef} contributes only to the vacuum loops
and will be dropped in the following.
The second term on the last line of \eqref{logkdef} has an expansion
in terms of Feynman diagrams.

Using this method and Eq.~\eqref{wptdef},
one can write down the 1-loop expression for the
vacuum polarisation tensor for the gauge bosons in a $U(N)$ noncommutative
gauge theory with massless fields, all in the adjoint representation, \cite{HKT},
\begin{align}
\Pi^{AB}_{\mu \nu}(k) =\, \hf \sum_{j} \alpha_{j} \int \! \frac{d^{d}p}{(2 \pi)^d} \,
&\left[ d(j)M^{AB}(k,p)\left(
\frac{1}{2}\frac{(2p+k)_\mu (2p+k)_\nu}{p^2(p+k)^2}
- \frac{\delta^{\mu \nu}}{p^2 } \right) \right. \nonumber \\
&\left. + 2 C(j) M^{AB}(k,p)\frac{\delta^{\mu \nu} - k_{\mu} k_{\nu}}
{p^2  (p+k)^2} \label{Pieqn}
\right]
\end{align}
where we have introduced the tensor
\EQ{
\label{emme}
M^{AB}(k, p) =
(-d^{ALM} \sin \frac{k \tilde{p}}{ 2}+ f^{ALM} \cos \frac{k \tilde{p}}{ 2})
(d^{BML} \sin \frac{k \tilde{p}}{2}+ f^{BML} \cos \frac{k \tilde{p}}{ 2})
\ \ .
}
To proceed further on, we rewrite  (\ref{emme}) using the relations
\cite{Bonora:2000ga}
\beqa
\label{tracce}
f^{ALM} f^{BML} &=& -N c_A \delta_{AB} \ \ ,
\nonumber \\
d^{ALM} d^{BML} &=& N d_A \delta_{AB} \ \ ,
\nonumber \\
f^{ALM} d^{BML} &=& 0 \ \ ,
\eeqa
where $c_A = 1 - \delta_{0A}$ and $d_A = 2-c_A$.
This way (\ref{emme}) collapses to
\EQ{
\label{mab}
M^{AB}(k,p) = - N \ \delta^{AB} (1-\delta_{0A}\cos \tilde{k}{p})
\ \ .
}
$\tilde{k}$ is defined as $\tilde{k}^\nu = k_\mu\theta^{\mu\nu}$.
Loop integrals involving the first term in equation \eqref{mab}
give rise to the planar contribution and are analogous to their commutative
counterparts. Integrals involving the second term in equation \eqref{mab}
give the non-planar contribution and cause the UV/IR mixing. Equation
\eqref{mab} already shows that it is exclusively degrees of freedom
associated with the generator $T^0 \propto \uno$ that will
exhibit the UV/IR mixing.

The sum in \eqref{Pieqn}
extends over all particles in the adjoint representation that appear in the loop (gauge fields,
ghosts, fermions and scalars). The constants $\alpha_j$, $d(j)$, and $C(j)$ are
as shown in Table (\ref{const_table}). Only matter in the adjoint representation
is considered here because matter in the fundamental representation does
not contribute to nonplanar diagrams.

Planar loop integrals are done in the dimensional regularisation, while nonplanar ones
are UV-finite and are calculated directly in 4 dimensions.
Nonplanar integrals are performed using
\beqa
\int \frac{d^4 p }{ (2\pi)^4}
\frac{e^{i p\tilde{k}}}{ p^2 (p+k)^2}&=&\frac{2 }{ (4\pi)^2} \int_{0}^{1}dx \
K_{0} (\sqrt{k^2x(1-x)} |\tilde{k}|)
\ \ ,
\\
\int \frac{d^4 p }{ (2\pi)^4}
\frac{e^{i p\tilde{k}}}{ p^2 }&=&
\frac{1}{(4\pi)^2}\frac{4}{ \tilde{k}^2}
\ \ ,
\eeqa
where the Bessel function $K_0(z)$ has a small-$z$ expansion
\EQ{
K_0 (z) = -\log \frac{z}{ 2}  \left[ 1 + \frac{z^2 }{ 4} + O(z^4)\right]
-\gamma_{E} - (\gamma_{E}-1)\frac{z^2 }{ 4} + O(z^4)
\ \ , \label{Bessel}
}
while for large-$z$ it is exponentially suppressed.
Using \eqref{pimunu} and \eqref{Pieqn}, planar contributions to
$\Pi_1^{AB}$ are given by
\EQ{
{\Pi_1^{AB}}_{\rm planar}=\, \frac{\delta^{AB}}{4 (4\pi)^2} \int_0^1 dx \,
b_0\,\log(k^2x(1-x)) \ ,
}
and nonplanar ones are
\EQ{
{\Pi_1^{AB}}_{\rm nonplanar}=\, \frac{\delta^{A0}\delta^{B0}}{4 (4\pi)^2} \int_0^1 dx \,
2 b_0\, K_0\left(\sqrt{k^2x(1-x)}|\tilde{k}|\right) \ ,
}
Now, using the definition of the running couplings \eqref{A.5}, and the
expansion \eqref{Bessel} we derive the asymptotic expressions \eqref{rg3},\eqref{rg1},\eqref{rg2}.

In order to generalise to $N_f\neq 0$ of fundamental flavours of matter fields,
we only need to recall that the fundamental fields in the loops do not contribute
to $\theta$-dependent diagrams, i.e. they contribute only to planar diagrams.
Hence,
\bea
{\Pi_1^{AB}}_{\rm planar} &=& \frac{\delta^{AB}}{4 (4\pi)^2} \int_0^1 dx \,
b_0^{N_f\neq 0}\,\log(k^2x(1-x)) \ , \\
{\Pi_1^{AB}}_{\rm nonplanar} &=& \frac{\delta^{A0}\delta^{B0}}{4 (4\pi)^2} \int_0^1 dx \,
2 b_0^{N_f=0}\, K_0\left(\sqrt{k^2x(1-x)}|\tilde{k}|\right) \ .
\eea

In order to include spontaneously broken gauge groups we need to add masses to
gauge bosons,
\begin{align}
\Pi^{AB}_{\mu \nu}(k) =\, \hf \sum_{j} \alpha_{j} \int \! \frac{d^{d}p}{(2 \pi)^d} \,
&\left[ d(j)M^{AB}(k,p)\left(
\frac{1}{2}\frac{(2p+k)_\mu (2p+k)_\nu}{(p^2  + m^2) [(p+k)^2+m^2]}
- \frac{\delta^{\mu \nu}}{p^2 + m^2} \right) \right. \nonumber \\
&\left. + 2 C(j) M^{AB}(k,p)\frac{\delta^{\mu \nu} - k_{\mu} k_{\nu}}
{(p^2  + m^2) [(p+k)^2+m^2]}
\right]
\end{align}
but it is important to note that the gauge boson masses depend on $L$ and $M$
indices in \eqref{emme}, and one should use \eqref{emme} rather than \eqref{mab}
for $M^{AB}$.
In order to proceed, we represent the final answer for $\Pi^{AB}_{\mu \nu}(k)$
as the contribution of the unbroken (massless) $U(N)$, plus the correction,
originating from
\SP{
\int d^4 p \left(
\frac{1}{(p^2  + m_L^2) [(p+k)^2+m_M^2]} -
\frac{1}{p^2(p+k)^2}\right) \\
\times \left(f^{ALM}f^{BML}\cos^2 \frac{\tilde{k}p}{ 2}-
d^{ALM}d^{BML} \sin^2 \frac{\tilde{k}p}{ 2}\right) \ .
}
The expression on the right hand side is UV-convergent in 4 dimensions, and one can set
$\tilde{k}=0$, i.e. remove the nonplanar effective UV cut-off
$\Lambda_{UV}=1/\tilde{k}$.
This is equivalent to saying that for
$\tilde{k}^2 \to 0$, the $\log(\tilde{k})$ terms will cancel in the integral.
What is left after that is the standard $\theta$-independent expression which
corrects the planar $SU(N)$ running from a massless to a massive case.
This has nothing to do with the $U(1)$ running coupling which is simply given by the
 massless contribution. From this reasoning,
Eqs.~\eqref{RGSU},\eqref{RGU1} follow.

\Appendix{Generators of SU(4)}

Throughout this paper,
the generators of SU(4) are $t^{a}=\frac{1}{2}T^{a}$ where $a=1..15$ and the
$T^{a}$ are:
\begin{align}
T_1=&\left(\begin{matrix}
0&1&0&0 \\
1&0&0&0 \\
0&0&0&0 \\
0&0&0&0 \\
\end{matrix}\right)&&
T_2=&\left(\begin{matrix}
0&-i&0&0 \\
i&0&0&0 \\
0&0&0&0 \\
0&0&0&0 \\
\end{matrix}\right)&&
T_3=&\left(\begin{matrix}
1&0&0&0 \\
0&-1&0&0 \\
0&0&0&0 \\
0&0&0&0 \\
\end{matrix}\right)
\nonumber
\\
T_4=&\left(\begin{matrix}
0&0&1&0 \\
0&0&0&0 \\
1&0&0&0 \\
0&0&0&0 \\
\end{matrix}\right)&&
T_5=&\left(\begin{matrix}
0&0&-i&0 \\
0&0&0&0 \\
i&0&0&0 \\
0&0&0&0 \\
\end{matrix}\right)&&
T_6=&\left(\begin{matrix}
0&0&0&0 \\
0&0&1&0 \\
0&1&0&0 \\
0&0&0&0 \\
\end{matrix}\right)
\nonumber
\\
T_7=&\left(\begin{matrix}
0&0&0&0 \\
0&0&-i&0 \\
0&i&0&0 \\
0&0&0&0 \\
\end{matrix}\right)&&
T_8=\frac{1}{\sqrt{3}}&\left(\begin{matrix}
1&0&0&0 \\
0&1&0&0 \\
0&0&-2&0 \\
0&0&0&0 \\
\end{matrix}\right)&&
T_9=&\left(\begin{matrix}
0&0&0&1 \\
0&0&0&0 \\
0&0&0&0 \\
1&0&0&0 \\
\end{matrix}\right)
\nonumber
\\
T_{10}=&\left(\begin{matrix}
0&0&0&i \\
0&0&0&0 \\
0&0&0&0 \\
-i&0&0&0 \\
\end{matrix}\right)&&
T_{11}=&\left(\begin{matrix}
0&0&0&0 \\
0&0&0&1 \\
0&0&0&0 \\
0&1&0&0 \\
\end{matrix}\right)&&
T_{12}=&\left(\begin{matrix}
0&0&0&0 \\
0&0&0&i \\
0&0&0&0 \\
0&-i&0&0 \\
\end{matrix}\right)
\nonumber
\\
T_{13}=&\left(\begin{matrix}
0&0&0&0 \\
0&0&0&0 \\
0&0&0&1 \\
0&0&1&0 \\
\end{matrix}\right)&&
T_{14}=&\left(\begin{matrix}
0&0&0&0 \\
0&0&0&0 \\
0&0&0&i \\
0&0&-i&0 \\
\end{matrix}\right)&&
T_{15}=\frac{1}{\sqrt{6}}&\left(\begin{matrix}
1&0&0&0 \\
0&1&0&0 \\
0&0&1&0 \\
0&0&0&-3 \\
\end{matrix}\right)
\nonumber
\end{align}
\newpage

\end{document}